\documentclass[
  aps,
  pra,
  amsmath,
  amssymb,
  reprint,
  superscriptaddress,
  floatfix,
  longbibliography
]{revtex4-2}

\usepackage[utf8]{inputenc}
\usepackage[T1]{fontenc}
\usepackage{graphicx}
\usepackage{dcolumn}
\usepackage{bm}
\usepackage{booktabs}
\usepackage{braket}
\usepackage{hyperref}

\graphicspath{{./}}

\newcommand{\authorcomment}[1]{}

\newcommand{\Tr}{\operatorname{Tr}}

\newcommand{\outerr}[2]{|{#1}\rangle\!\langle{#2}|}

\hypersetup{
  colorlinks=true,
  linkcolor=blue,
  citecolor=blue,
  urlcolor=blue,
  pdftitle={Exact Quasiprobability Hierarchy of the Double-Morse Oscillator},
  pdfauthor={F. Chogle, B. Teklu, and M. F. Pereira},
  pdfsubject={Manuscript prepared for Physical Review A}
}
\begin{document}
\raggedbottom

\title[Exact Quasiprobability Hierarchy]%
{Exact Cahill–Glauber Hierarchy for the Double-Morse Ground State}

\author{F. Chogle}
\affiliation{%
College of Computing and Mathematical Sciences,
Department of Applied Mathematics and Sciences,
Khalifa University of Science and Technology,
Abu Dhabi 127788, United Arab Emirates}

\author{B. Teklu}
\affiliation{%
College of Computing and Mathematical Sciences,
Department of Applied Mathematics and Sciences,
Khalifa University of Science and Technology,
Abu Dhabi 127788, United Arab Emirates}
\affiliation{%
KU Research Center for Intelligent Computing, Networks and Sensing (ICONS), Khalifa University of Science and Technology, 127788, Abu Dhabi, United Arab Emirates}
\email{berihu.gebrehiwot@ku.ac.ae}

\author{M. F. Pereira}
\affiliation{%
Department of Physics,
Khalifa University of Science and Technology,
Abu Dhabi 127788, United Arab Emirates}
\affiliation{%
Institute of Physics of the Czech Academy of Sciences,
Prague, Czech Republic}
\email{mauro.pereira@ku.ac.ae}\date{August 22, 2026}

\begin{abstract}
\noindent Phase-space and quasiprobability methods now play operational roles in
quantum technologies, characterizing localization, non-Gaussianity,
nonclassical resources, and coarse-graining. We develop an exact,
representation-consistent analysis of the lowest quasi-exact ground
state of the symmetric double-Morse oscillator. In the double-Morse
potential, the dimensionless parameter $A$ controls the separation of
the minima and the central barrier, thereby changing the physical
ground state. At fixed $A$, the Cahill--Glauber parameter $s$ labels the
quasiprobability $W_A^{(s)}(q,p)$: $s=0$, $-1$, and $1$ give the
Wigner, Husimi $Q$, and Glauber--Sudarshan $P$ representations,
respectively. Although the potential is double-welled for $0<A<1$, the
exact ground-state amplitude is single-peaked at the origin and lies
above the barrier; as $A$ approaches unity, the merged well remains
locally quartic rather than harmonic. Closed analytical expressions are
obtained for the Wigner function and Weyl characteristic function. The
Wigner function displays the $A$-dependent exchange between position
and momentum localization and retains negative regions, certifying
nonclassicality and, for this pure state, non-Gaussianity. The Weyl
function is its Fourier dual, generates symmetrically ordered moments
and cumulants, and yields the full $s$-ordered hierarchy. For $s<0$,
isotropic Gaussian smoothing suppresses fine sign-changing structure
while preserving the large-scale localization envelope. The Husimi
endpoint is nonnegative without implying classicality, whereas the
$P$ representation remains distributional. Thus, $A$ controls the
physical phase-space geometry, while $s$ controls how the same
non-Gaussian and nonclassical state is resolved across complementary
representations.
\end{abstract}

\maketitle
\section{Introduction}

Phase-space and quasiprobability methods have acquired operational roles
across quantum technologies, extending well beyond their original use
as visual descriptions of quantum states. An early application to
open-system quantum control showed that continuous homodyne monitoring
and feedback can be used to preserve and manipulate the coherence of an
atomic two-level system, with measurement backaction represented as a
quantum diffusion of the Bloch vector
\cite{PhysRevA.57.4877}. In odd-dimensional discrete-variable
systems, negativity of a discrete Wigner representation identifies
resources relevant to magic-state quantum computation and helps
distinguish processes that remain amenable to efficient classical
simulation \cite{Veitch_2012}. In constrained qubit-computation
schemes for which the allowed measurements preserve Wigner
nonnegativity, contextuality and Wigner-function negativity are
likewise necessary resources and are closely related to the breakdown
of efficient classical simulation
\cite{PhysRevA.95.052334}. Phase-space methods are also being
developed for quantum machine learning, where the
Stratonovich--Weyl correspondence allows multi-qubit states and
dynamics to be represented by functions on symplectic manifolds and
provides a basis for variational modelling directly in phase space
\cite{heightman2025}.

In continuous-variable, optical, atomic, and mechanical platforms,
phase-space methods similarly support the preparation, reconstruction,
and verification of non-Gaussian quantum states. Controlled
Wigner-negative multimode optical states, multiphonon mechanical
excitations, massive mechanical Schr\"odinger-cat states, and highly
squeezed bosonic code states have all been investigated through
phase-space diagnostics
\cite{RaEtAl2020,ChuEtAl2018,BildEtAl2023,KendellEtAl2024}.
Quasiprobability distributions have also been used to characterize
nonlinearity, information-theoretic quantities, and nonclassical
behavior in nanomechanical and ultracold systems
\cite{Teklu2015Nonlinearity,Ughradar2026}.

These applications exploit different properties of phase-space
representations. Negativity can reveal quantum resources that are
incompatible with an ordinary classical probability model; the geometry
of a distribution records localization, squeezing, and interference;
characteristic functions provide access to moments and
displacement-dependent correlations; and coherent-state
representations distinguish coarse-grained localization from the
existence of a classical statistical mixture of coherent states. This
diversity of roles motivates treating quasiprobabilities as a
complementary hierarchy rather than selecting one representation solely
as a visualization tool.

Because position and momentum do not commute, a quantum state cannot in
general be represented by an ordinary positive joint probability
density on classical phase space. Wigner introduced a real and
normalized quasiprobability whose marginals reproduce the position- and
momentum-space probability densities \cite{wigner1932}. Subsequent
developments established phase-space methods as a complete
operator--symbol calculus
\cite{Moyal1949,AgarwalWolf1970,Hillery1984,Ferrie2011}. The full
Wigner function is therefore informationally complete despite being
potentially negative. Its Fourier transform, the Weyl characteristic
function, contains the same state information reorganized into
displacement-dependent overlaps, moments, and cumulants.

The Wigner representation is the symmetrically ordered member of the
Cahill--Glauber $s$-parameterized family
\cite{CahillGlauber1969,GerryKnight2023}. At $s=0$, one obtains the
Wigner function. At the antinormally ordered endpoint $s=-1$, Gaussian
smoothing produces the Husimi $Q$ function, a nonnegative distribution
with a coherent-state-overlap interpretation
\cite{Husimi1940,CahillGlauber1969}. At the normally ordered endpoint
$s=1$, the Glauber--Sudarshan $P$ representation asks whether the
density operator can be expressed as a classical statistical mixture
of coherent states
\cite{Glauber1963,Sudarshan1963,Cahill1965}. Depending on the state and
the chosen ordering, the same density operator may thus be represented
by a regular signed function, a smooth positive density, or a
generalized distribution.

\begingroup
Different orderings emphasize complementary physical questions:
the survival of structure under Gaussian coarse-graining, Wigner-level
negativity, and the existence of a positive coherent-state mixture
\cite{Hillery1984,Lee1991,KieselVogel2010,Ferrie2011,HughesEtAl2014}.
\par\endgroup

Nonlinear and anharmonic oscillators provide a natural setting in which
to study these distinctions. Nonquadratic potentials can produce
non-Gaussian eigenstates and generate oscillatory Wigner structure and
negativity even when the position-space wavefunction is not an explicit
superposition of separated packets. Ground-state non-Gaussianity has
therefore been used to quantify oscillator nonlinearity, while
anharmonic dynamics and nonlinear transformations have been
investigated as resources for producing nonclassical states
\cite{ParisEtAl2014,AlbarelliEtAl2016,OlivaSteuernagel2019,
RosiekEtAl2024,MooreFilip2025}.
Complementary driven-dissipative approaches show that phase-space
geometry and quadrature correlations can also be engineered through
coherent driving, reservoir design, and parametric control. In a
non-Hermitian degenerate parametric oscillator, coherent driving
combined with a squeezed reservoir can enhance both intracavity and
output quadrature squeezing \cite{Wodedo2026Environment}. In
nanomechanical resonators, parametric amplification combined with
two-tone laser control can enhance two-mode mechanical squeezing while
retaining robustness against thermal noise
\cite{Wodedo2025Amplifying}. These studies concern engineered squeezing
rather than Wigner negativity itself, but they reinforce the broader
point that Hamiltonian nonlinearity, external driving, and dissipation
can all reshape phase-space localization and correlations.
Semiconductor materials and devices provide a particularly relevant
extension of this phase-space perspective. Wigner-function transport
formulations have been applied to resonant-tunnelling structures,
metal--oxide--semiconductor field-effect transistors, and biased
semiconductor superlattices, including partially coherent and
dissipative carrier transport in GaAs/AlGaAs systems relevant to
quantum-cascade devices \cite{weinbub2018recent}. In the optical
domain, Wigner--Weyl methods combined with localization-landscape
theory have been used to describe absorption in disordered
semiconductor alloys \cite{banon2022wigner}. In parallel, exactly
solvable Hamiltonians have provided compact analytical frameworks for
relating semiconductor luminescence and its temperature dependence to
material parameters and sample quality \cite{ma11010002}.
Taken together, these developments motivate analytically controlled
benchmark models in which the effects of Hamiltonian nonlinearity,
driving, dissipation, disorder, and device geometry can ultimately be
distinguished from changes introduced solely by the chosen phase-space
representation. Such a separation is particularly valuable when
extending quasiprobability methods from idealized nonlinear oscillators
toward semiconductor transport and optical-response problems.

Despite their common mathematical origin, the Wigner, Weyl, Husimi, and
Glauber--Sudarshan representations of nonlinear oscillator states are
often considered separately. This can obscure a fundamental
distinction. A change in a Hamiltonian parameter modifies the physical
state, whereas a change in the ordering parameter modifies only the
representation and phase-space resolution with which that state is
described. A unified analysis of an analytically known state across the
complete hierarchy is therefore needed to separate physical changes in
phase-space geometry from changes caused by Gaussian smoothing or
normal ordering.

The symmetric double-Morse oscillator provides an analytically
tractable platform for such an analysis. Its dimensionless parameter
$A$ controls the geometry of the potential, including the separation of
the minima and the height of the internal barrier. Previous work has
shown that the double-Morse system supports tunable non-Gaussianity,
Wigner negativity, entanglement, and metrological behavior
\cite{Chogle2026}. Related studies of wide double-well and nonharmonic
potentials have demonstrated the generation of delocalized motional
states and complex Wigner structure, providing a dynamical counterpart
to the stationary ground state considered here
\cite{RodaLlordesEtAl2024,RieraCampenyEtAl2024}. The quasi-exact
solvability of the double-Morse model is particularly useful because it
allows the effect of a physical control parameter to be propagated
analytically through several complementary phase-space
representations.

In the present work, we develop an exact and
representation-consistent phase-space description of the lowest
quasi-exact double-Morse ground state. We derive closed analytical
expressions for the
Wigner function and the Weyl characteristic function, use the latter to
generate Weyl-symmetrized moments and cumulants, and construct the
associated $s$-ordered hierarchy. The Husimi and
Glauber--Sudarshan endpoints are then examined as, respectively, a
positive coherent-state-overlap distribution and a generally
distributional test of coherent-state classicality.

The novelty of the work is not merely the calculation of several
quasiprobability functions. The exact solution shows that,
although the potential is double-welled for $0<A<1$, the lowest
quasi-exact ground-state amplitude is single-peaked at the origin and
lies above the internal barrier. Moreover, when the two minima
coalesce as $A\to1^{-}$, the resulting well remains locally quartic
rather than becoming harmonic. The state therefore retains a
higher-than-quadratic, non-Gaussian structure throughout the parameter
range considered. The combined analysis follows how these non-Gaussian
features appear
across the ordering hierarchy as $A$ varies.

The Weyl formulation also connects the exact analysis to
experimentally oriented methods based on characteristic functions and
moments, which can witness Wigner negativity and non-Gaussian
correlations without relying solely on a full reconstruction of the
quasiprobability distribution \cite{Zaw2024,MallickEtAl2025}.

The paper is organized as follows. Section~\ref{sec:DMmodel}
introduces the double-Morse potential, derives the lowest quasi-exact
ground state, and analyzes its position- and momentum-space structure.
Section~\ref{sec:wignerFunc} derives the exact Wigner function and Weyl
characteristic function and develops their moment and cumulant
interpretation. Section~\ref{sec:cahillGlauberHierarchy} constructs the
Cahill--Glauber hierarchy, examines Gaussian smoothing and the Husimi
and Glauber--Sudarshan endpoints, and compares physical control by $A$
with representational control by $s$. Section~\ref{sec:Conclusion}
summarizes the principal conclusions and outlines quantitative
extensions.

\begin{figure*}[htbt]
  \centering
  \includegraphics[width=\textwidth]{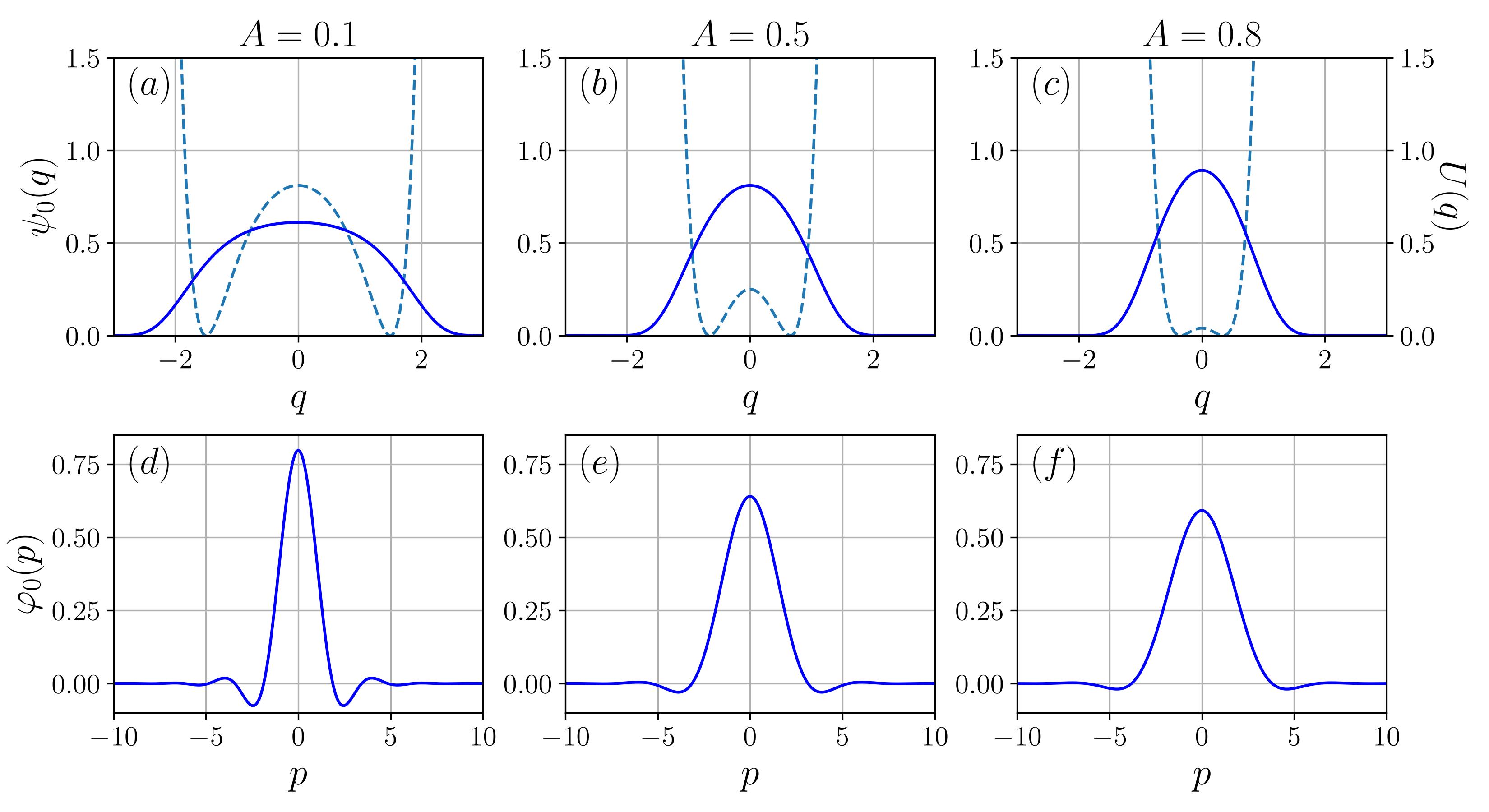}
  \caption{Ground-state position- and momentum-space wavefunctions of the
  quasi-exact double-Morse oscillator for $\mu=1$. Top row, panels \mbox{(a)--(c)}:
  position-space amplitude $\psi_0(q)$ from Eq.~\eqref{eq:psi0}, shown as a
  solid curve, together with the dimensionless potential
  \mbox{$U(q)=[A\cosh(2q)-1]^2$}, shown as a dashed curve. Bottom row, panels (d)--(f):
  momentum-space amplitude $\varphi_0(p)$ from Eq.~\eqref{eq:phi0}. Columns
  correspond, from left to right, to $A=0.1$, $0.5$, and $0.8$. The
  dimensionless coordinates are $q=\alpha x/2$ and
  $p=2P_x/(\hbar\alpha)$. As $A$ increases, $\psi_0(q)$ narrows, whereas
  $\varphi_0(p)$ broadens and its oscillatory side lobes weaken. For
  $0<A<1$, the potential is double-welled, while the $n=0$ quasi-exact
  ground-state amplitude remains even and single-peaked at $q=0$.}
  \label{fig:groundStateWavefunctions}
\end{figure*}

\section{Double-Morse Model: Quasi-Exact Solution}\label{sec:DMmodel}

The symmetric double-Morse (DM) potential is
\begin{equation}
  V(x)=D\left[A\cosh(\alpha x)-1\right]^2,
  \label{eq:dmPotential}
\end{equation}
where $D$ sets the overall energy scale, $\alpha^{-1}$ sets the characteristic length scale, and the dimensionless parameter $A$ controls both the separation of the minima and the height of the central barrier. All three parameters are positive. For $0<A<1$, the potential has a double-well structure.

Consider a particle of mass $m$ in the DM potential. Its position-space wavefunction $\psi(x)$ obeys the stationary Schr\"odinger equation
\begin{equation}
  -\frac{\hbar^2}{2m}\frac{d^2\psi(x)}{dx^2}+V(x)\psi(x)=E\psi(x),
  \label{eqn:dmSE}
\end{equation}
where $E$ is the energy. Introducing the dimensionless variables
\begin{equation}
  q=\frac{\alpha x}{2},\qquad
  \mu^2=\frac{8mD}{\hbar^2\alpha^2},\qquad
  \epsilon=\frac{8mE}{\hbar^2\alpha^2},
  \label{eq:dimensionlessTransform}
\end{equation}
gives
\begin{equation}
  \frac{d^2\psi(q)}{dq^2}+\left[\epsilon-U(q)\right]\psi(q)=0,
  \label{eq:dimensionlessSE}
\end{equation}
where the dimensionless potential is
\begin{equation}
  U(q)=\mu^2\left[A\cosh(2q)-1\right]^2.
  \label{eq:DimensionlessU}
\end{equation}

The normalizable asymptotic factor of Eq.~\eqref{eq:dimensionlessSE} is
$\exp[-\mu A\cosh(2q)/2]$. We therefore use the ansatz
\begin{equation}
  \psi(q)=\exp\left[-\frac{\mu A}{2}\cosh(2q)\right]\phi(q),
  \label{eq:ansatz}
\end{equation}
for which the auxiliary function satisfies
\begin{equation}
  \frac{d^2\phi}{dq^2}
  -2A\mu\sinh(2q)\frac{d\phi}{dq}
  +\left[\widetilde{\epsilon}
  +2A\mu(\mu-1)\cosh(2q)\right]\phi=0,
  \label{eq:phiODE}
\end{equation}
with $\widetilde{\epsilon}=\epsilon-\mu^2(1+A^2)$.

\begingroup
The quasi-exact condition $\mu=n+1$, $n\in\mathbb N_0$, for Eq.~\eqref{eq:phiODE} defines an infinite discrete family of Hamiltonians. For each fixed $n$, however, the corresponding invariant sector has finite dimension $n+1$. The integer $n$ labels this sector and the associated Hamiltonian parameter; it is not the excitation index of a fixed potential. The auxiliary functions in that sector have the form $\phi(q)=e^{-nq}P_n(e^{2q})$, with $P_n$ of degree at most $n$. The full confining Hamiltonian has infinitely many bound states, of which this construction determines only a finite subset. Higher sectors permit analogous phase-space calculations for their algebraic eigenstates; excited states outside the algebraic sector require a separate spectral calculation. The double-Morse potential thus belongs to the class of quasi-exactly solvable potentials \cite{Razavy1980,KonwentEtAl1998,FinkelEtAl1999,UshveridzeBook}.
\par\endgroup

The lowest quasi-exact sector has $n=0$, so the condition $\mu=n+1$ fixes $\mu=1$. Equation~\eqref{eq:phiODE} then admits the constant solution $\phi(q)=1$, while $\widetilde{\epsilon}=0$. The resulting normalizable, node-free eigenfunction is the ground state, with
\begin{align}
  \psi_0(q)
  &=\frac{1}{\sqrt{K_0(A)}}
    \exp\left[-\frac{A}{2}\cosh(2q)\right],
  \label{eq:psi0}\\
  \epsilon_0&=1+A^2.
  \label{eqn:e0}
\end{align}
Here $K_\nu$ is the modified Bessel function of the second kind. The normalization coefficient follows from
\begin{equation}
  K_\nu(z)=\int_0^\infty dt\,e^{-z\cosh(t)}\cosh(\nu t).
  \label{eq:KvInt}
\end{equation}
Thus, in the calculations below, $A$ is varied while the quasi-exact condition $\mu=1$ is held fixed. For fixed $m$ and $\alpha$, this condition is equivalent to $D=\hbar^2\alpha^2/(8m)$.

The ground-state momentum-space wavefunction is obtained by Fourier transforming Eq.~\eqref{eq:psi0} according to
\begin{equation}
  \varphi(p)=\frac{1}{\sqrt{2\pi}}
  \int_{-\infty}^{\infty}dq\,\psi(q)e^{-ipq}.
  \label{eq:generalFT}
\end{equation}
The momentum coordinate $p$, like $q$, is dimensionless; the corresponding physical momentum is $P_x=(\hbar\alpha/2)p$. Using the evenness of $\psi_0(q)$, the substitution $t=2q$, and Eq.~\eqref{eq:KvInt}, one obtains
\begin{align}
  \varphi_0(p)
  &=\frac{1}{\sqrt{2\pi K_0(A)}}
    \int_0^\infty dt\,
    e^{-\frac{A}{2}\cosh(t)}
    \cos\left(\frac{pt}{2}\right)
    \notag\\
  &=\frac{1}{\sqrt{2\pi K_0(A)}}K_{ip/2}\left(\frac{A}{2}\right).
  \label{eq:phi0}
\end{align}
Because $\psi_0(q)$ is real and even, $\varphi_0(p)$ is also real and even. For small $A$, the broad position-space amplitude produces a narrow central momentum-space peak accompanied by rapidly decaying, sign-changing side lobes. As $A$ increases, the position wavefunction narrows and the momentum wavefunction broadens. This reciprocal behavior follows from Fourier duality, while the side lobes reflect the non-Gaussian spatial profile.

In the double-well regime $0<A<1$, the minima of the dimensionless potential are located at
\begin{equation}
  q_{\mathrm{min}}
  =\pm\frac{1}{2}\operatorname{arcosh}\left(\frac{1}{A}\right),
  \label{eq:qmin}
\end{equation}
where $U(q_{\mathrm{min}})=0$. In the $\mu=1$ sector considered here, the dimensionless central barrier height is
\begin{equation}
  B=U(0)-U(q_{\mathrm{min}})=(1-A)^2.
  \label{eq:barrier}
\end{equation}
As $A\to1^-$, the two minima approach the origin and the central barrier vanishes. The coalescence of the minima does not, however, produce a harmonic-oscillator limit. At $A=1$,
$U(q)=[\cosh(2q)-1]^2=4q^4+\mathcal{O}(q^6)$ near the origin, so the merged single well remains anharmonic.

Before turning to the quasiprobability analysis, it is useful to identify the configuration-space origin of the phase-space structures discussed below. From Eq.~\eqref{eq:psi0},
$\psi_0'(q)=-A\sinh(2q)\psi_0(q)$. Because $\psi_0(q)>0$, its only stationary point is $q=0$, where $\psi_0''(0)=-2A\psi_0(0)<0$. The exact ground-state amplitude therefore has a unique maximum at the origin, even though the potential has two distinct minima for $0<A<1$.

In addition, Eqs.~\eqref{eqn:e0} and \eqref{eq:barrier} give
\begin{equation}
  \epsilon_0-B=(1+A^2)-(1-A)^2=2A>0,
  \label{eq:sum}
\end{equation}
throughout the double-well regime. The exact state therefore lies above the internal barrier and should not be interpreted as a conventional below-barrier tunnelling doublet. For small $A$, the wavefunction extends broadly across both wells and the intervening barrier region rather than separating into two well-localized components.

\begingroup
The inequality $\epsilon_0-B=2A>0$ applies specifically to the $\mu=1$ sector. Below-barrier states occur in higher sectors. For example, at $\mu=2$ the exact even ground state is proportional to $e^{-A\cosh(2q)}\cosh q$, with energy $3+4A^2-4A$. Since the barrier is $4(1-A)^2$, this state lies below it when $0<A<1/4$. The odd first excited state is proportional to $e^{-A\cosh(2q)}\sinh q$, with energy $3+4A^2+4A$, and is also below the barrier for $0<A<1/12$. Their phase-space transforms can be evaluated by extending the same integral construction to the hyperbolic prefactors. Superpositions of the pair provide a route to coherent tunnelling dynamics, whereas a stationary eigenstate alone does not exhibit population transfer.
\par\endgroup

The role of $A$ can also be seen directly from the exact envelope. Near the origin,
\begin{equation}
  \begin{aligned}
    \psi_0(q)
    &=\frac{1}{\sqrt{K_0(A)}}
      \exp\left[-\frac{A}{2}\cosh(2q)\right]\\
    &=\frac{e^{-A/2}}{\sqrt{K_0(A)}}
      \exp\left[-Aq^2-\frac{A}{3}q^4
      +\mathcal{O}(q^6)\right].
  \end{aligned}
  \label{eq:expansion}
\end{equation}
The quadratic term determines the leading Gaussian scale of the central peak, while the quartic and higher even powers encode departures from Gaussian form generated by the anharmonic potential. Increasing $A$ contracts the position-space distribution, whereas Fourier duality produces the corresponding broadening in momentum space. The oscillatory momentum-space side lobes at small $A$ therefore originate from the extended, non-Gaussian spatial profile and should not be attributed to interference between two separated position-space packets.

Figure~\ref{fig:groundStateWavefunctions} thus anticipates two features that reappear throughout the subsequent phase-space analysis: an $A$-dependent exchange between position and momentum localization, and non-Gaussian corrections that cannot be described by a purely quadratic oscillator model.

\begin{figure*}[t]
  \centering
  \includegraphics[width=\textwidth]{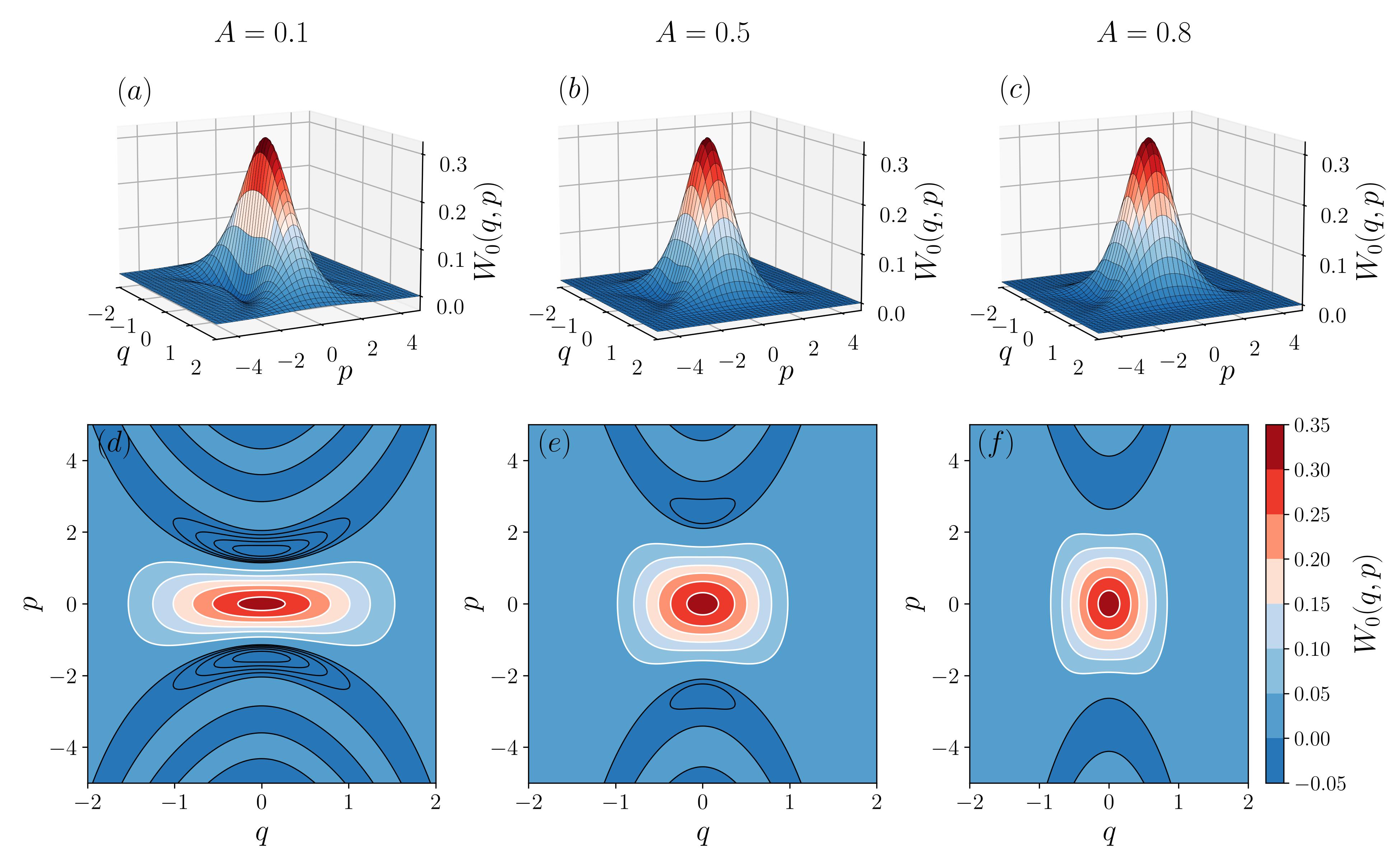}
  \caption{Three-dimensional surface plots (top row, panels (a)--(c))
  and signed two-dimensional contour maps (bottom row, panels (d)--(f)) of
  the ground-state Wigner function $W_0(q,p)$, Eq.~\eqref{eq:w0}, for
  $\mu=1$. Columns correspond, from left to right, to $A=0.1$, $0.5$, and
  $0.8$. Both rows represent the full signed Wigner function. Owing to the
  viewing angle and the comparatively small vertical extent of the negative
  lobes, the negative regions are less visually prominent in the surface
  plots and are more readily identified in the contour maps. White contours
  denote positive levels, while black contours denote zero and negative
  levels. As $A$ increases, the central positive region narrows along $q$ and
  broadens along $p$, while the visible negative lobes become less extensive
  but remain present.}
  \label{fig:wignerCombined}
\end{figure*}

\section{Wigner distribution and Weyl characteristic function}
\label{sec:wignerFunc}

\subsection{Ground-state Wigner function}
\label{subsec:wignerGroundState}

For a quantum state described by the density operator $\hat{\rho}$, we use the Wigner-function convention
\begin{equation}
  W(q,p)
  = \frac{1}{\pi}\int_{-\infty}^{\infty}\!dy\,
  \braket{q+y|\hat{\rho}|q-y}e^{-2ipy}.
  \label{eq:wignerGen}
\end{equation}
For a pure state $\hat{\rho}=\outerr{\psi}{\psi}$, Eq.~\eqref{eq:wignerGen} becomes
\begin{equation}
  W(q,p)
  = \frac{1}{\pi}\int_{-\infty}^{\infty}\!dy\,
  \psi^{\ast}(q+y)\psi(q-y)e^{-2ipy}.
  \label{eqn:wignerPure}
\end{equation}
Although $W(q,p)$ need not be nonnegative, it is normalized,
\begin{equation*}
  \int_{-\infty}^{\infty}\!dq
  \int_{-\infty}^{\infty}\!dp\,W(q,p)=1,
\end{equation*}
and its marginals reproduce the position- and momentum-space probability densities:

\begin{equation}
\begin{aligned}
  \rho_q(q) &\equiv \int_{-\infty}^{\infty}\!dp\,W(q,p)
  = |\psi(q)|^2,\\
  \rho_p(p) &\equiv \int_{-\infty}^{\infty}\!dq\,W(q,p)
  = |\varphi(p)|^2.
\end{aligned}
\label{eq:marginals}
\end{equation}
More generally, the inverse Weyl transform establishes a one-to-one correspondence between $W(q,p)$ and $\hat{\rho}$, so the full Wigner function is informationally complete despite its possible negativity \cite{Hillery1984,Ferrie2011}.

For the exact DM ground state, the product entering Eq.~\eqref{eqn:wignerPure} is
\begin{equation*}
  \psi_0^\ast(q+y)\psi_0(q-y)
  = \frac{1}{K_0(A)}
  \exp\!\left[-A\cosh(2q)\cosh(2y)\right],
\end{equation*}
where we used
$\cosh[2(q+y)]+\cosh[2(q-y)]=2\cosh(2q)\cosh(2y)$.
Substitution into Eq.~\eqref{eqn:wignerPure}, followed by $t=2y$ and use of Eq.~\eqref{eq:KvInt}, gives the exact result
\begin{equation}
  W_0(q,p)
  = \frac{K_{ip}\!\left(A\cosh(2q)\right)}{\pi K_0(A)}.
  \label{eq:w0}
\end{equation}
For real $p$ and positive argument, $K_{ip}$ is real and even in $p$. Equation~\eqref{eq:w0} therefore shows that $W_0$ is real and even in both phase-space coordinates. It also yields
\begin{align*}
  W_0(0,0) &= \frac{1}{\pi},\\
  W_0(q,0) &= \frac{K_0\!\left(A\cosh(2q)\right)}{\pi K_0(A)}>0.
\end{align*}
Thus the central value is independent of $A$, while sign changes occur away from the $p=0$ axis.
The changing orientation of the central Wigner structure is the phase-space counterpart of the wavefunction trends in Fig.~\ref{fig:groundStateWavefunctions}. Because the marginals in Eq.~\eqref{eq:marginals} reproduce the position and momentum probability densities, the broad horizontal core at small $A$ reflects a state that is extended in $q$ and localized in $p$. As $A$ increases, the core contracts along $q$ and expands along $p$, becoming predominantly vertical.

The sign-changing lobes contain information beyond this large-scale width geometry. In Eq.~\eqref{eqn:wignerPure}, points symmetrically separated about $q$ contribute through the nonlocal product $\psi^{\ast}(q+y)\psi(q-y)$. For the present state, the resulting oscillations are not fringes between two independently localized packets, because the exact wavefunction is single-peaked and lies above the internal barrier. Instead, they arise from the non-Gaussian wavefunction generated by the anharmonic potential.

For a pure state, Hudson's theorem states that an everywhere nonnegative Wigner function is possible only for a Gaussian state \cite{Hudson1974}. The negative regions in Fig.~\ref{fig:wignerCombined} therefore witness both nonclassicality and non-Gaussianity of the present ground state \cite{AlbarelliEtAl2018}. As $A$ increases, the wavefunction becomes more concentrated around the origin and the visible negative lobes become less extensive in the three cases shown. Their reduced visual extent should not, however, be interpreted as a transition to a classical or exactly Gaussian state. Throughout $0<A<1$, the exact pure state retains higher-than-quadratic terms and remains non-Gaussian; consequently, the exact Wigner function must retain some negative region. The three displayed cases demonstrate a redistribution of the visible negative structure, but they do not by themselves establish monotonic behavior of a quantitative measure such as the Wigner-negative volume.

Related nonlinear-oscillator models, including self-Kerr and nonlinear nanomechanical systems, likewise exhibit Wigner fringes and negative regions under nonlinear evolution or engineered steady-state preparation \cite{AlbarelliEtAl2016,StobinskaEtAl2008,RipsEtAl2012,RosiekEtAl2024,RieraCampenyEtAl2024,MooreFilip2025}.

\begin{figure*}[t]
  \centering
  \includegraphics[width=\textwidth]{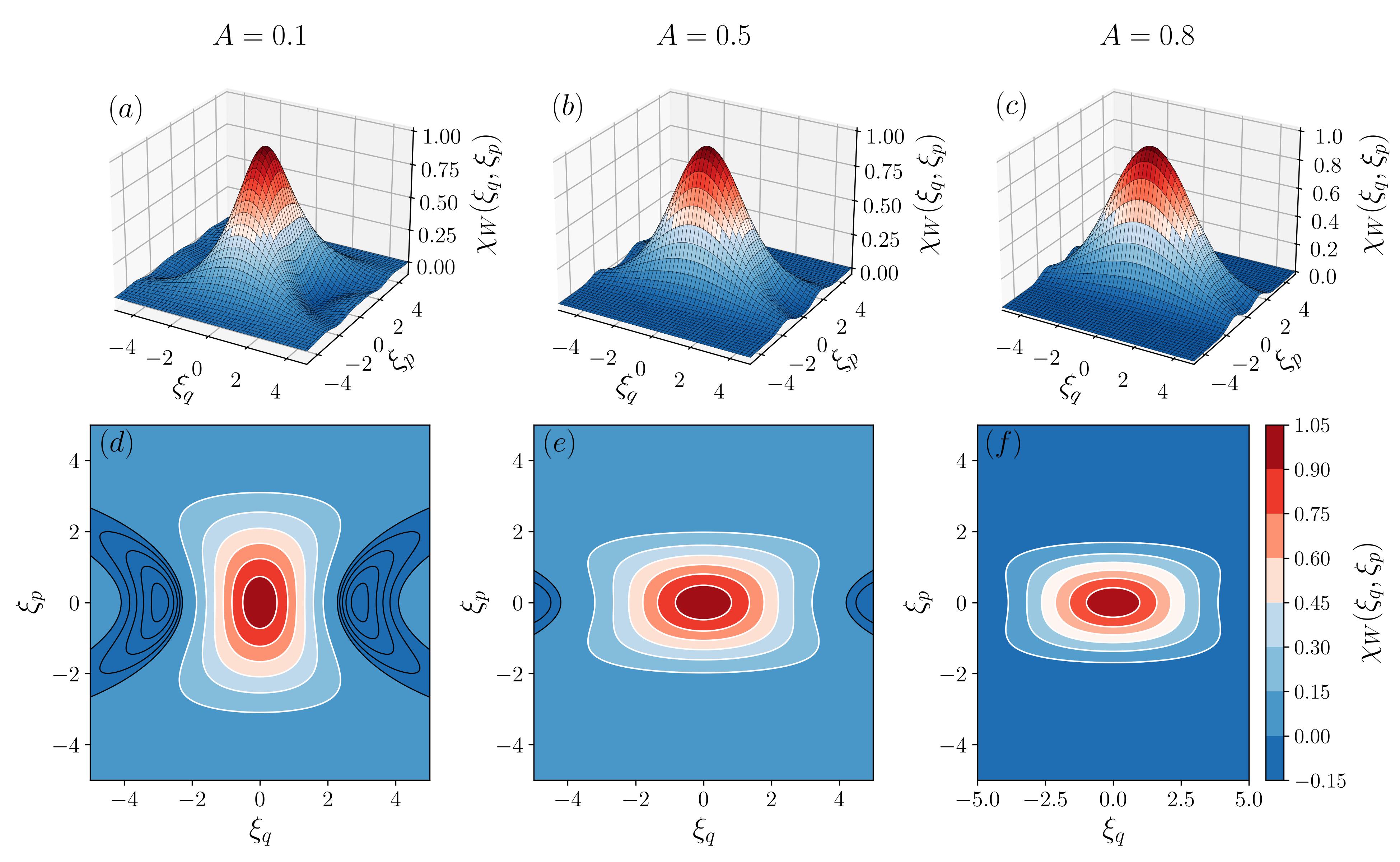}
  \caption{Three-dimensional surface plots (top row, panels (a)--(c))
  and signed two-dimensional contour maps (bottom row, panels (d)--(f)) of
  the exact Weyl characteristic function $\chi_W(\xi_q,\xi_p)$,
  Eq.~\eqref{eq:chiW0}, for the quasi-exact double-Morse ground state with
  $\mu=1$. Columns correspond, from left to right, to $A=0.1$, $0.5$, and
  $0.8$. Both rows represent the full signed characteristic function. Owing
  to the viewing angle and the comparatively small vertical extent of the
  negative side lobes, the negative regions are less visually prominent in
  the surface plots and are more readily identified in the contour maps. The
  function is real and even in both variables and satisfies
  $\chi_W(0,0)=1$. White contours denote positive levels, while black
  contours denote zero and negative levels. As $A$ increases, the central
  lobe broadens along $\xi_q$, narrows along $\xi_p$, and the sign-changing
  side lobes become less prominent.}
  \label{fig:weylCharacteristic}
\end{figure*}

\subsection{Weyl characteristic function and moments}
\label{subsec:weylMoments}

The same complete state information can be reorganized in the Fourier domain through the Weyl, or symmetrically ordered, characteristic function
\begin{align}
  \chi_W(\xi_q,\xi_p)
  &= \Tr\!\left[\hat{\rho}\,
     e^{i(\xi_q\hat{q}+\xi_p\hat{p})}\right] \notag\\
  &= \int_{-\infty}^{\infty}\!dq
     \int_{-\infty}^{\infty}\!dp\,
     W(q,p)e^{i(\xi_q q+\xi_p p)}.
  \label{eq:chiWGen}
\end{align}
Performing the $p$ integration with Eq.~\eqref{eqn:wignerPure} gives the equivalent wavefunction-overlap form
\begin{equation*}
  \chi_W(\xi_q,\xi_p)
  = \int_{-\infty}^{\infty}\!dq\,
    \psi_0^\ast\!\left(q+\frac{\xi_p}{2}\right)
    \psi_0\!\left(q-\frac{\xi_p}{2}\right)
    e^{i\xi_q q}.
\end{equation*}
Here $\xi_q$ is Fourier conjugate to $q$, while $\xi_p$ controls the separation between the two wavefunction arguments. Substituting Eq.~\eqref{eq:psi0}, using the same hyperbolic addition identity as above, and applying Eq.~\eqref{eq:KvInt} yield
\begin{equation}
  \chi_W(\xi_q,\xi_p)
  = \frac{K_{i\xi_q/2}\!\left(A\cosh\xi_p\right)}{K_0(A)}.
  \label{eq:chiW0}
\end{equation}
Equation~\eqref{eq:chiW0} is normalized by $\chi_W(0,0)=1$ and is real and even in both Fourier variables. Its two axial slices are the ordinary characteristic functions of the position and momentum marginals:
\begin{equation}
  \begin{aligned}
    \chi_W(\xi_q,0)
    &= \int_{-\infty}^{\infty}\!dq\,
       \rho_q(q)e^{i\xi_q q}
     = \frac{K_{i\xi_q/2}(A)}{K_0(A)},\\
    \chi_W(0,\xi_p)
    &= \int_{-\infty}^{\infty}\!dp\,
       \rho_p(p)e^{i\xi_p p}
     = \frac{K_0\!\left(A\cosh\xi_p\right)}{K_0(A)}.
  \end{aligned}
  \label{eq:chiWxiq}
\end{equation}
The second slice is positive and decreases with $|\xi_p|$, whereas the first may develop sign-changing side lobes through the imaginary order of the Bessel function. A sign change of $\chi_W$ is a Fourier-domain feature and should not be identified directly with Wigner negativity.

Figure~\ref{fig:weylCharacteristic} displays the Fourier-domain
counterpart of the phase-space structure shown in
Fig.~\ref{fig:wignerCombined}. The axial slices in
Eq.~\eqref{eq:chiWxiq} also connect directly to the position- and
momentum-space distributions in Fig.~\ref{fig:groundStateWavefunctions}.
A broad distribution in a physical variable produces a narrow
characteristic function in its conjugate Fourier variable. Thus, at
smaller $A$, the broader position distribution produces a narrower
central profile along $\xi_q$, while the more localized momentum
distribution produces a broader profile along $\xi_p$. As $A$ increases,
the position distribution contracts and the momentum distribution
broadens; correspondingly, $\chi_W$ broadens along $\xi_q$ and narrows
along $\xi_p$. This reciprocal deformation provides an independent
consistency check on the position-space, momentum-space, and Wigner
results. The sign-changing side lobes along $\xi_q$ provide a
Fourier-domain signature of the non-Gaussian structure generated by the
anharmonic potential.

The full two-variable characteristic function generates all Weyl-symmetrized moments:
\begin{equation*}
  \left\langle
    \{\hat{q}^{m}\hat{p}^{n}\}_{\mathrm{sym}}
  \right\rangle
  = (-i)^{m+n}
    \left.
    \frac{\partial^{m+n}\chi_W}
         {\partial\xi_q^{m}\partial\xi_p^{n}}
    \right|_{\xi_q=\xi_p=0}.
\end{equation*}
Because the ground-state Wigner function is even in both $q$ and $p$, every moment that is odd in either variable vanishes. In particular, the nonzero position moments are
\begin{equation}
  \left\langle \hat{q}^{2n}\right\rangle
  = (-1)^n
    \left.
    \frac{\partial^{2n}}{\partial\xi_q^{2n}}
    \chi_W(\xi_q,0)
    \right|_{\xi_q=0},
  \label{eq:qnMoment}
\end{equation}
and the momentum moments are
\begin{equation}
  \left\langle \hat{p}^{2n}\right\rangle
  = (-1)^n
    \left.
    \frac{\partial^{2n}}{\partial\xi_p^{2n}}
    \chi_W(0,\xi_p)
    \right|_{\xi_p=0}.
  \label{eq:pnMoment}
\end{equation}
For example, the exact second moments are
\begin{align*}
  \left\langle \hat{q}^{2}\right\rangle
  &= \frac{1}{4K_0(A)}
     \left.
     \frac{\partial^2 K_{\nu}(A)}{\partial\nu^2}
     \right|_{\nu=0},\\
  \left\langle \hat{p}^{2}\right\rangle
  &= A\frac{K_1(A)}{K_0(A)}.
\end{align*}
Since the first moments vanish, these are also the position and momentum variances. They provide a quantitative counterpart to the position--momentum width exchange seen in the figures.

In a neighborhood of the origin, where $\chi_W$ is nonzero, the
same information can be organized in terms of symmetrically ordered
phase-space cumulants:
\begin{equation}
  \ln\chi_W(\xi_q,\xi_p)
  = \sum_{\substack{m,n\geq 0\\m+n\geq 1}}
    \frac{\kappa_{mn}}{m!\,n!}
    (i\xi_q)^m(i\xi_p)^n .
  \label{eq:weylCumulants}
\end{equation}
Here $\kappa_{mn}$ denotes the Weyl-symmetrized cumulant of order
$(m,n)$. Because the state and its Wigner function are even in both
$q$ and $p$, every cumulant that is odd in either index vanishes.

For the present centered state, the second-order cumulants coincide
with the variances,
\begin{equation}
  \kappa_{20}=\langle\hat q^2\rangle,
  \qquad
  \kappa_{02}=\langle\hat p^2\rangle.
\end{equation}
The leading marginal cumulants that probe non-Gaussianity are
\begin{equation}
  \kappa_{40}
  =\langle\hat q^4\rangle
   -3\langle\hat q^2\rangle^2,
  \qquad
  \kappa_{04}
  =\langle\hat p^4\rangle
   -3\langle\hat p^2\rangle^2.
\end{equation}
For a Gaussian state, $\ln\chi_W$ is a polynomial of total order at
most two; hence all cumulants of total order greater than two vanish.
Nonzero higher-order cumulants therefore provide a direct diagnostic
of non-Gaussianity \cite{WeedbrookEtAl2012}.

Higher-order cumulants alone do not generally imply Wigner negativity
for an arbitrary mixed state. Nevertheless, suitably constructed
characteristic-function and moment-based witnesses can certify Wigner
negativity or non-Gaussian correlations without requiring complete
phase-space reconstruction \cite{Zaw2024,MallickEtAl2025}. For the
pure state considered here, non-Gaussianity also implies Wigner
negativity by Hudson's theorem \cite{Hudson1974}.

The Weyl characteristic function also generates the complete
Cahill--Glauber hierarchy: multiplication by an ordering-dependent
Gaussian factor produces the ordered characteristic functions introduced
in the following section.

\begin{figure*}[t]
  \centering
  \includegraphics[width=\textwidth]{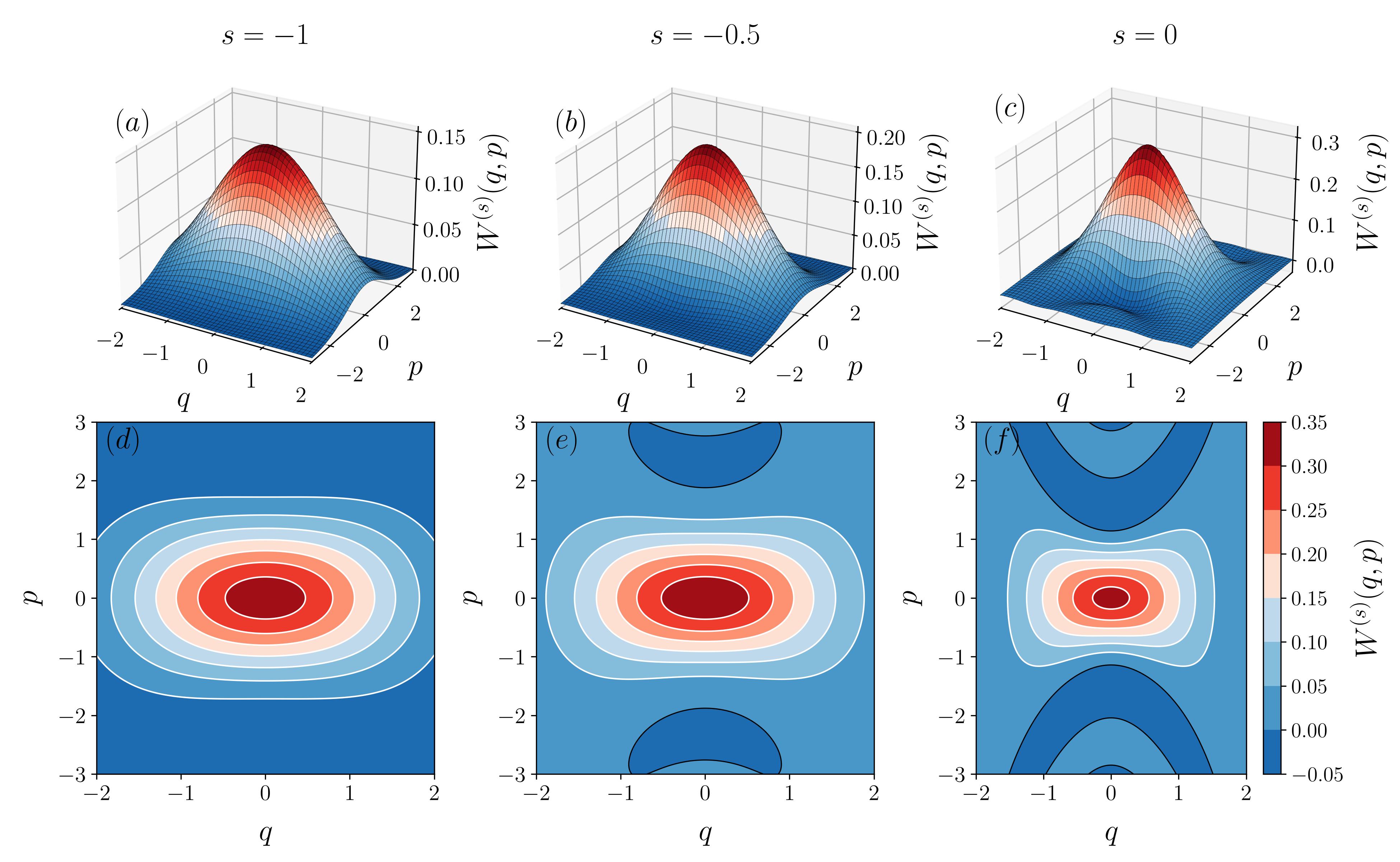}
  \caption{Three-dimensional surface plots (top row, panels (a)--(c))
  and signed two-dimensional contour maps (bottom row, panels (d)--(f)) of
  the Cahill--Glauber $s$-ordered quasiprobability distributions
  $W^{(s)}(q,p)$ for the quasi-exact double-Morse ground state with $\mu=1$
  and $A=0.1$. From left to right, $s=-1$, $-0.5$, and $0$, corresponding to
  the Husimi endpoint $W^{(-1)}$, an intermediate Gaussian-smoothed
  distribution, and the Wigner function $W^{(0)}=W_0$, respectively. Negative regions are more readily identified in the contour maps than in the corresponding three-dimensional surface plots. White contours denote positive levels, while black contours denote the zero and negative levels. The top-row vertical ranges are panel dependent,
  whereas the lower row uses a common signed color scale. Moving from $s=0$
  toward $s=-1$, Gaussian smoothing broadens the central structure and
  progressively suppresses the negative lobes, yielding a nonnegative
  distribution at $s=-1$.}
  \label{fig:sOrderedPlots}
\end{figure*}

\section{Cahill-Glauber Hierarchy}\label{sec:cahillGlauberHierarchy}
\subsection{Ordered characteristic functions}\label{subsec:sFamily}
\begingroup
Cahill and Glauber introduced the one-parameter ($s$-parameter) family of phase-space characteristic functions that unifies the standard orderings \cite{CahillGlauber1969,Hillery1984,Ferrie2011}.
\endgroup

The representation of the $s$-parameter is given by 
\begin{equation}
  \chi^{(s)}(\xi_q,\xi_p) = e^{s(\xi_q^2 + \xi_p^2)/4}\chi_W(\xi_q,\xi_p)
  \label{eq:chiSGen}
\end{equation}
where $\chi_W$ is the Weyl characteristic function introduced in the previous section. 
In the present comparison, the ordering parameter is restricted to $-1\leq s\leq 1$.
At the extremes, $s=1$ and $s=-1$, $\chi^{(s)}$ correspond to normal and antinormal ordered characteristic functions.   
The inverse Fourier transform of Eq.~\eqref{eq:chiSGen} gives the associated quasiprobability distribution

\begin{align}
  W^{(s)}(q,p)
  &= \frac{1}{(2\pi)^2}
  \int_{-\infty}^{\infty}\!d\xi_q
  \int_{-\infty}^{\infty}\!d\xi_p\,
  \chi^{(s)}(\xi_q,\xi_p) \notag\\
  &\quad\times e^{-i(\xi_q q + \xi_p p)}.
  \label{eq:Ws}
\end{align}
When $s=1,0$ and $-1$, the $P$-function, the Wigner function, and the Husimi $Q$-function are obtained using Eq.~\eqref{eq:Ws}.
For $s<0$, Eq.~\eqref{eq:Ws} represents a stable Gaussian smoothing of the Wigner function:
\begin{align}
  W^{(s)}(q,p) &= \frac{1}{\pi(-s)} \int_{-\infty}^{\infty}\int_{-\infty}^{\infty} dq'dp'\, W(q',p') \notag \\  
  &\times \exp{\left(-\frac{(q-q')^2 + (p-p')^2}{(-s)}\right)} \label{eq:WsGaussian}. 
\end{align}

This ordering flow also underlies nonclassical-depth and filtered-quasiprobability constructions: the amount or type of smoothing required to obtain a regular nonnegative distribution provides a representation-aware nonclassicality diagnostic \cite{Lee1991,KieselVogel2010}.

Figure~\ref{fig:sOrderedPlots} shows the ordering flow at fixed
$A=0.1$. Moving from $s=0$ toward $s=-1$, from right to left in the
figure, applies the Gaussian smoothing in Eq.~\eqref{eq:WsGaussian}
to the same density operator; it does not describe a time evolution.
Two effects are visible during this smoothing flow. First, the negative
lobes decrease in magnitude and extent and ultimately disappear at the
Husimi endpoint. Second, the predominantly $q$-extended orientation of
the central phase-space envelope survives. Because the smoothing kernel
in Eq.~\eqref{eq:WsGaussian} is isotropic, it broadens the distribution
but does not generate or rotate this anisotropy. The unequal extensions
along $q$ and $p$ are inherited from the position and momentum widths of
the underlying state at $A=0.1$. By contrast, the sign-changing regions
are finer-resolution features and are therefore more sensitive to
Gaussian coarse-graining.

At $s=-1$, the distribution is the Husimi representation and is
nonnegative by construction. This positivity should not be interpreted
as evidence that the underlying state has become classical. The density
operator is unchanged as $s$ varies, and the same state that possesses
Wigner-negative regions at $s=0$ is represented by a positive,
Gaussian-smoothed distribution at $s=-1$. The comparison therefore
shows that Wigner negativity and coarse-grained phase-space localization
answer different questions: the Wigner function exposes sign-sensitive
non-Gaussian structure, whereas the Husimi representation retains the
large-scale localization envelope at coherent-state resolution.

The three selected values $s=0,-0.5$, and $-1$ illustrate this ordering
flow but do not determine a numerical nonclassical depth. Such a
determination would require scanning $s$ continuously, within the stated
$s$-parameter convention, and locating the threshold value at which
$W^{(s)}(q,p)$ first becomes everywhere nonnegative.

\subsection{Husimi \texorpdfstring{$Q$}{Q}-function}\label{subsec:Qfunction}

\begin{figure*}[t]
    \centering
    \includegraphics[width=\textwidth]{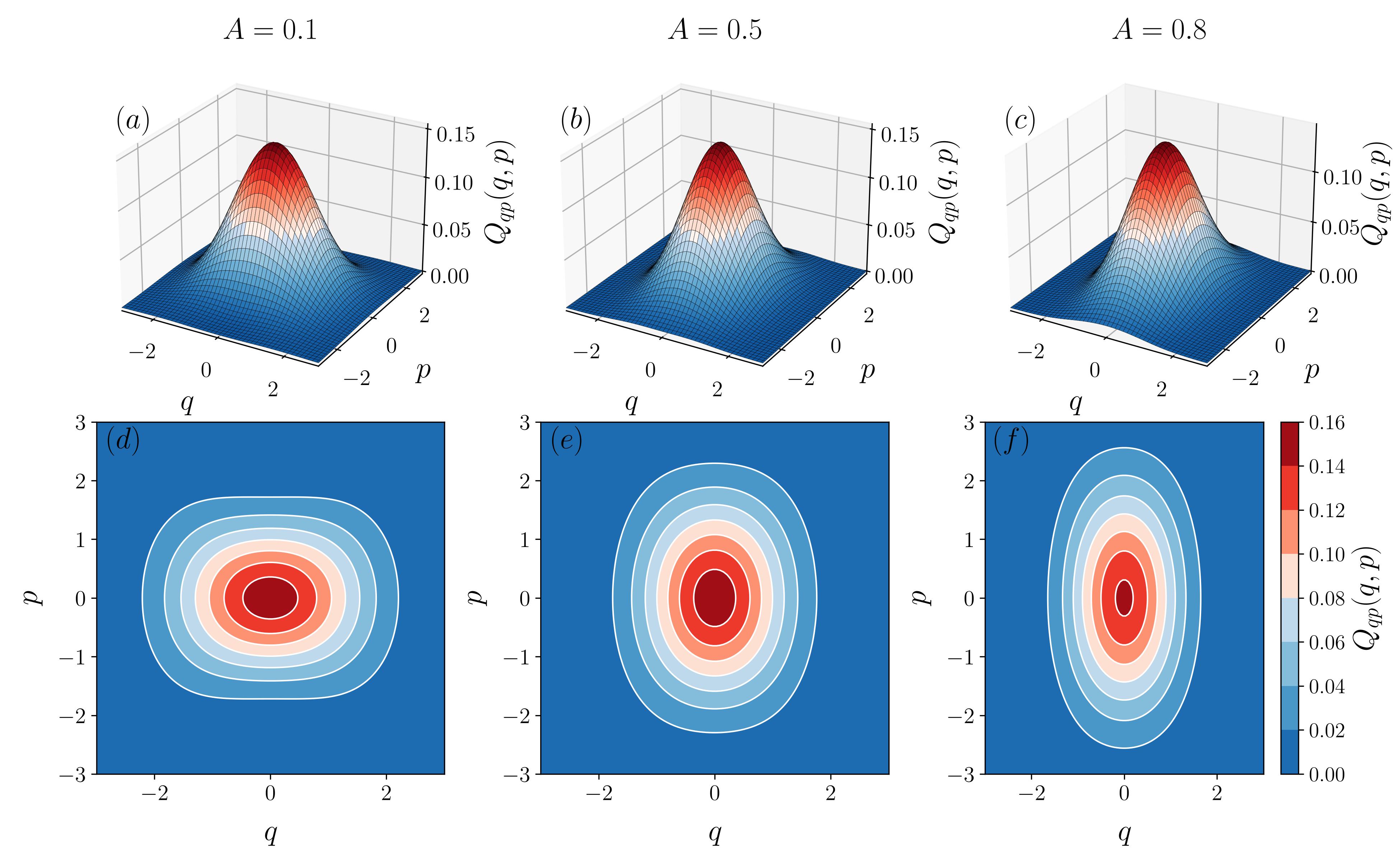}
  
  \caption{Three-dimensional surface plots (top row, panels (a)--(c))
  and two-dimensional contour maps (bottom row, panels (d)--(f)) of the
  $q,p$-normalized Husimi distribution
  $Q_{qp}(q,p)\equiv \tfrac{1}{2}Q\!\left((q+ip)/\sqrt{2}\right)
  =W^{(-1)}(q,p)$ for the quasi-exact double-Morse ground state with $\mu=1$.
  Columns correspond, from left to right, to $A=0.1$, $0.5$, and $0.8$.
  Both rows display the same nonnegative distributions. As $A$ increases, the
  distribution narrows along $q$ and broadens along $p$, displaying in
  Gaussian-smoothed form the position--momentum localization trade-off. The
  lower row uses a common nonnegative color scale, whereas the top-row
  vertical ranges are panel dependent.}
    \label{fig:Qfunction}
\end{figure*}

In coherent state representation, the Husimi $Q$-function is the diagonal element of the density operator, given by \cite{Husimi1940,CahillGlauber1969}
\begin{equation}
  Q(\beta) = \frac{\braket{\beta|\hat{\rho}|\beta}}{\pi}.
  \label{eq:QfuncFormula}
\end{equation}
For comparison with the $s$-ordered distributions expressed in the
$(q,p)$ variables, we use the $q,p$-normalized Husimi density
\begin{equation*}
  Q_{qp}(q,p)
  \equiv
  \frac{1}{2}
  Q\!\left(\frac{q+ip}{\sqrt{2}}\right)
  =
  W^{(-1)}(q,p).
\end{equation*}
The factor $1/2$ follows from
$d^2\beta=dq\,dp/2$ and ensures that
\begin{equation*}
  \int_{-\infty}^{\infty}dq
  \int_{-\infty}^{\infty}dp\,
  Q_{qp}(q,p)=1.
\end{equation*}

Expanding the ground state in the number-state basis gives
$\ket{\psi_0}=\sum_n C_n\ket{n}$, with $C_n=\braket{n|\psi_0}$.
Substituting this into Eq.~\eqref{eq:QfuncFormula} yields

\begin{equation}
  Q(\beta)
  =
  \frac{e^{-|\beta|^2}}{\pi}
  \left|
    \sum_{n=0}^{\infty}
    C_n\frac{(\beta^\ast)^n}{\sqrt{n!}}
  \right|^2.
  \label{eq:QfuncBeta}
\end{equation}

Although $Q$ is nonnegative because it is a Gaussian-smoothed phase-space density, it remains sensitive to non-Gaussian structure; for a single mode, the zeros of the Husimi function organize the stellar hierarchy of non-Gaussian states \cite{ChabaudEtAl2020}.

Figure~\ref{fig:Qfunction} fixes the ordering at $s=-1$ and varies
$A$, complementing the fixed-$A$ comparison in
Fig.~\ref{fig:sOrderedPlots}. The $A=0.1$ panels in the two figures
represent the same distribution, $Q_{qp}(q,p)=W^{(-1)}(q,p)$.
At small $A$, the Husimi density is extended predominantly along $q$
and localized along $p$. As $A$ increases, it contracts along $q$ and
broadens along $p$. Comparison with Fig.~\ref{fig:wignerCombined}
shows that this localization geometry remains visible at coherent-state
resolution after Gaussian smoothing removes the sign-changing lobes.

Through Eq.~\eqref{eq:QfuncFormula}, the Husimi density identifies
regions of appreciable overlap with coherent states. Whether the state
can itself be expressed as a positive statistical mixture of coherent
states is determined by the $P$ representation considered next.

\subsection{Glauber--Sudarshan \texorpdfstring{$P$}{P}-representation}\label{subsec:Pfunction}

The defining diagonal representation is
\begin{equation}
  \hat{\rho}=\int_{\mathbb{C}} d^2\beta\,P(\beta)\ket{\beta}\!\bra{\beta},
  \label{eq:Pdiagonal}
\end{equation}
and a regular nonnegative $P$ corresponds to a classical statistical mixture of coherent states \cite{Sudarshan1963,Cahill1965}.
The $P$-distribution function can be obtained through the $s$-parameterized family introduced by Cahill and Glauber.
For the present non-Gaussian pure state, however, the $s=1$ inverse transform is not an ordinary integrable function and must be interpreted distributionally \cite{Lee1991}.
Another way to obtain the $P$-function is the coherent-state inversion due to Mehta \cite{Mehta1967}.
The corresponding inversion formula is
\begin{equation}
  P(\beta) = \frac{e^{|\beta|^2}}{\pi^2} \int_\mathbb{C} d^2u\, e^{|u|^2}\braket{-u|\hat{\rho}|u} e^{u^\ast\beta - u\beta^\ast}
  \label{eq:PfuncGen}
\end{equation}

where $\ket{u}$ is a coherent state.
Substituting the number state expansion in Eq.~\eqref{eq:PfuncGen} yields,
\begin{equation}
  P(\beta) = e^{|\beta|^2}\sum_{n,m}\frac{C_n}{\sqrt{n!}}\frac{C_m^\ast}{\sqrt{m!}}
  \left(-\frac{\partial}{\partial\beta}\right)^n\left(-\frac{\partial}{\partial\beta^\ast}\right)^m \delta^{(2)}(\beta)
  \label{eq:Pfunction}
\end{equation}
where $\delta^{(2)}(\beta)$ is the 2D Dirac $\delta$-function and the summation is understood to be from $0$ to $\infty$.
Since the expression contains derivatives of the $\delta$-function, it should be read distributionally.

It does not define an ordinary surface, and the mere appearance of delta derivatives is not a quantitative ``degree of singularity'': even a coherent state has a delta-distributed $P$ representation. A pure state can possess a nonnegative $P$ probability measure only if it is itself a coherent state, because any nontrivial positive mixture of distinct coherent states is mixed \cite{Cahill1965}. The state in Eq.~\eqref{eq:psi0}  is non-Gaussian and hence noncoherent, so it is $P$-nonclassical for every finite $A>0$.

A regularized $P$ plot is meaningful only after defining a filter. For example, with $\chi_N\equiv\chi^{(1)}$, one may replace $\chi_N$ by $\chi_N\Omega_w$ with a stated nonclassicality filter $\Omega_w$ and then demonstrate convergence as the width $w$ changes~\cite{Lee1991,KieselVogel2010}. The filter and its width therefore form part of the specification of any regularized visualization.

\subsection{State control and complementarity of the phase-space
representations}
\label{subsec:complementarity}

The parameter $A$ changes the double-Morse potential and its ground
state, whereas $s$ selects the ordering used to represent a fixed density
operator. These two controls organize the results as
\begin{equation*}
  A
  \longrightarrow
  U_A(q)
  \longrightarrow
  \ket{\psi_A}
  \longrightarrow
  \hat{\rho}_A,
  \qquad
  (A,s)\longmapsto W_A^{(s)}(q,p).
\end{equation*}
Figures~\ref{fig:wignerCombined} and~\ref{fig:Qfunction} vary $A$
at fixed $s=0$ and $s=-1$, respectively, while
Fig.~\ref{fig:sOrderedPlots} varies $s$ at fixed $A=0.1$.
Figure~\ref{fig:weylCharacteristic} gives the corresponding
Fourier-domain description as $A$ changes.

Read together, these comparisons distinguish the localization geometry
shared by the Wigner and Husimi distributions from the finer
sign-changing structure resolved at Wigner ordering. The moments and
cumulants in Sec.~\ref{subsec:weylMoments} provide quantitative
descriptors of these features, while the coherent-state-mixture
criterion in Sec.~\ref{subsec:Pfunction} establishes nonclassicality.
Moving toward positive $s$ reverses Gaussian smoothing and amplifies
fine Fourier-domain structure, as expressed by Eq.~\eqref{eq:chiSGen}.
The hierarchy thus relates localization, non-Gaussianity, and
coherent-state classicality within a single analytic framework.

\section{Conclusion}
\label{sec:Conclusion}

We have established an exact, representation-consistent
Cahill-Glauber hierarchy for the lowest quasi-exact double-Morse
ground state at $\mu=1$, using closed analytical expressions for
the Wigner function and Weyl characteristic function. For
$0<A<1$, the ground-state amplitude remains single-peaked and
lies above the central barrier despite the double-well potential.
As $A\to1^{-}$, the minima merge into a locally quartic rather
than harmonic well, and the state remains non-Gaussian.
 
The hierarchy distinguishes changes in the physical state,
controlled by $A$, from changes in operator ordering, controlled
by $s$. Increasing $A$ contracts the position distribution and
broadens its momentum counterpart, while Wigner negativity
reveals nonclassicality. The Weyl characteristic function
generates moments, cumulants, and the ordered family. Husimi
positivity does not imply classicality; the Glauber-Sudarshan
representation instead tests whether a positive coherent-state
mixture exists.
 
Quantitative extensions include evaluating the Wigner-negative
volume as a function of $A$, determining the nonclassical depth,
and constructing filtered $P$ quasiprobabilities with specified
filtering and convergence procedures. Higher quasi-exact sectors
provide access to algebraic excited states and below-barrier
regimes, with superpositions of the $\mu=2$ doublet offering a
route to coherent tunnelling. Longer-term directions include
adapting the hierarchy as a benchmark for semiconductor transport
and optical-response models before incorporating device-specific
physics.
 
The exact stationary state also provides an initial condition
for quenched, driven, or open-system dynamics. Closed-system
Wigner evolution follows the Moyal equation, whereas a specified
Lindblad model describes Markovian dissipation. Exact
characteristic-function propagation for quadratic Hamiltonians
with linear Lindblad operators~\cite{FernandesNeto2023} offers
a useful comparison, but does not imply exact solvability for
the nonquadratic double-Morse potential. Numerical and
controlled semiclassical approaches provide natural extensions.
The Cahill--Glauber relation generates the hierarchy at each
time; changing $s$ alone does not describe physical evolution.


\begin{acknowledgments}
This research was funded by Khalifa University of
Science and Technology through the Project ID: KU-INT-RIG-2024-8474000739.
\end{acknowledgments}

\section*{Author Contributions}

M.F.P. conceived the original project. All authors contributed to the derivation of the analytical results. F.C. and B.T. carried out the numerical implementation and generated all figures. B.T. secured the funding for the project. All authors contributed to the analysis and interpretation of the results and to the writing and revision of the manuscript.

\section*{Data Availability}
The data used to generate the plots in this manuscript were obtained by direct implementation of the equations and parameters provided in the text. The corresponding data can be obtained from the corresponding author upon reasonable request.

\bibliography{references_updated}

\end{document}